\documentclass[%
 reprint,
showpacs,preprintnumbers,
 amsmath,amssymb,
 aps,
]{revtex4-1}

\usepackage{graphicx}
\usepackage{dcolumn}
\usepackage{bm}

\usepackage{amssymb}
\usepackage{amsmath}
\usepackage{color}
\usepackage{graphicx}

\usepackage{dcolumn}
\usepackage{bm}
\usepackage{hyperref}
\usepackage{setspace}
\usepackage{bbm}
\usepackage{MnSymbol}

\usepackage{times}
\usepackage{mathdots}
\usepackage{mathrsfs}

\usepackage{setspace}

\usepackage{mathdots}
\usepackage{mathrsfs}
\usepackage{changes}

\begin{document}

\title{Topological properties of a generalized spin-orbit-coupled Su-Schrieffer-Heeger model}
\author{Masoud Bahari}
\author{Mir Vahid Hosseini}
 \email[Corresponding author: ]{mv.hosseini@znu.ac.ir}
\affiliation{Department of Physics, Faculty of Science, University of Zanjan, Zanjan 45371-38791, Iran}

\begin{abstract}
We explore theoretically the effect of inter and intra cell spin-orbit couplings on topological properties of a generalized Su-Schrieffer-Heeger model with multipartite lattice structure containing even number of sites per unit cell. We show that the spin-orbit couplings enrich topological phase diagrams so that nontrivial topological regions in the space of parameters extend significantly which is suitable for potential applications. We present an analytical formula for winding number in terms of sublattice number signaling that there are two nontrivial topological phases in the space of parameters hosting twofold/fourfold degenerate zero-mode boundary states. We also find that by increasing the number of sublattices within unit cell, the nontrivial regions of topological phase diagram including twofold degenerate zero-mode edge states extend dramatically. Our symmetry investigation shows that U(1) spin rotational symmetry around the lattice direction is the crucial ingredient for the appearance of fourfold degenerate zero-mode boundary states.
\end{abstract}

\maketitle

\section{Introduction}
The vast majority of attention among researchers has been focused on topological phases of matters \cite{Review1,Review2,Review3,Review4,Review6}. Also, a large amount of theoretical and experimental work has been carried out owing to potential applications of topological states as perfect conducting symmetry protected boundary states \cite{Ex1,Ex2,Ex3,theo1,theo2} in fault-tolerant quantum computation. There have been several models to realize topological insulators/superconductors in two and three dimensions \cite{Review1,Review2}. Also, one-dimensional (1D) systems, for instance, Su-Schrieffer-Heeger (SSH) chain \cite{SSH1} in the presence of sublattice and/or spin degrees of freedom are capable of exhibiting nontrivial phases providing topological insulators \cite{Chain2,Chain5,Chain6,Chain7,Chain8,Chain9,Chain10}. In this framework, several theoretical works have been dedicated to investigating effects of spin-orbit coupling \cite{SO}, Zeeman magnetic field \cite{Zeeman,SOZeeman}, and curvature \cite{Curvature} on topological properties of 1D superlattices by characterizing phase transition points \cite{Phase transition}. In addition, a new type of artificial quantum matter lattice composed of multilayer heterostructures of topological insulators and ordinary insulators has been fabricated experimentally which is condensed matter realization of SSH model \cite{Artificial}.

\indent Recently, it has been shown that 1D spin-orbit coupled nano-wires containing odd number of sublattices can be served as topological metal platforms \cite{topometal}. However, a little attention has been paid to systems including even number of sublattices with more than two sublattices per unit cell \cite{MoreSub1,MoreSub2} in the presence of spin-orbit coupling. In this context, two important and interesting questions have been remained unanswered: Is it possible to get to a wide range of topologically nontrivial phases via spin-orbit coupling with an enlarged unit cell? More specific, what is the effect of spin degree of freedom on the topological properties of a 1D superlattice containing even number of sublattices per unit cell?

\indent In this work, we generalize SSH model to cases with enlarged unit cells subjected to spin-orbit couplings. We show that in the absence of spin-orbit couplings, there are topologically trivial and nontrivial phases supporting none and twofold-degenerate zero-mode boundary states, respectively. On the contrary, in the presence of spin-orbit coupling, the topological phase diagrams undergo drastic changes so that the region including nontrivial phase is extended giving rise to appearance of two distinct nontrivial topological phases hosting twofold or fourfold-degenerate zero-mode boundary states. The system exhibits particle-hole, chiral, reflection, and effective time-reversal symmetries belonging to BDI topological class. Furthermore, for any even number of sublattices, the relevant topological invariant is determined analytically by winding number calculation through the bulk properties of quantum states. This, subsequently, unveils that topological region of one pair of edge states expands in the phase diagram with increasing the number of sublattices.
\begin{figure}[t]
\begin{center}
\includegraphics[width=8.5cm]{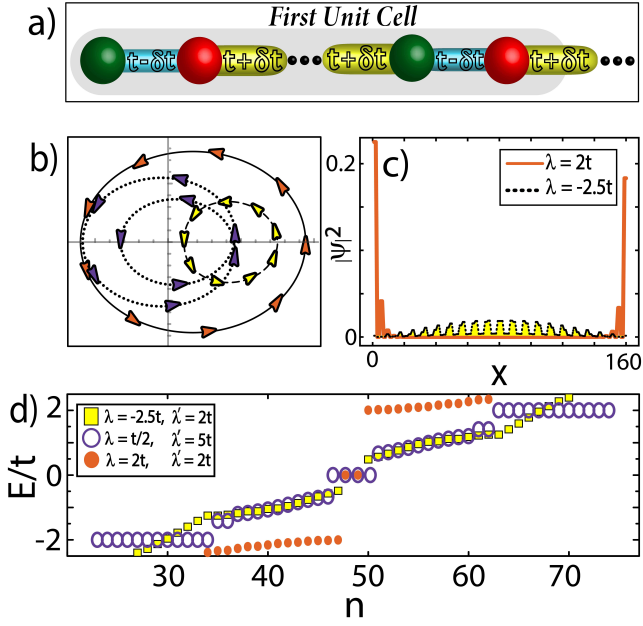}
\caption{(Color online) (a) Schematic illustration of 1D multipartite lattice structure having even number of sublattices per unit cell. (b) Parametric plot of $Re(Det\ \hat{F}_k)$ and $Im(Det\ \hat{F}_k)$ for $T=4$ and $\delta t=t/2$.  The dashed (solid) [(dotted)] line with yellow (orange) [(purple)] arrow corresponds to $\lambda=-2.5t$ and $\lambda^{\prime}=2t$ ($\lambda=\lambda^{\prime}=2t$) [($\lambda=t/2$ and $\lambda^{\prime}=5t$)].
(c) Probability distribution for two different values $\lambda=2t$ and $\lambda=-2.5t$. Here, $N=160$, $\delta t=t/2$ and $\lambda^{\prime}=2t$. (d) The energy spectrum of the system for $N=96$ versus wave function index corresponding to the panel (b) with the same parameters.
\label{Energy spectrum}}
\end{center}
\end{figure}

\section{Model and Theory}
\label{sec:1}
As shown in Fig. \ref{Energy spectrum} (a), we consider a 1D multipartite superlattice containing even number of sites per unit cell which is subjected to inter and intra unit cell spin-orbit couplings. The total tight binding reflection symmetric Hamiltonian $\hat{H}$ of the system has two terms as $\hat{H} = \hat{H}_{t}+\hat{H}_{s},$ where $\hat{H}_t$ and $H_{s}$ correspond to the kinetic Hamiltonian and spin-orbit one, respectively, given by
\begin{eqnarray}\label{hamiltonian-real space1}
\hat{H}_{t}&=&\sum_{n,\sigma}\sum_{\alpha=1}^{T}[t_{\alpha}\hat{c}^\dagger_{\alpha,n,\sigma}\hat{c}_{\alpha+1,n,\sigma}+h.c.],\\
\hat{H}_{so}&=&\sum_{n,\sigma}\sum_{\alpha=1}^{T}[\lambda_\alpha(-1)^{\alpha+1}\hat{c}^\dagger_{\alpha,n,\sigma}\hat{c}_{\alpha+1,n,-\sigma}+h.c.],
\end{eqnarray}
where $\hat{c}^\dagger_{\alpha,n,\sigma} (\hat{c}_{\alpha,n,\sigma})$  is the fermion creation (annihilation) operator of electron with spin $\sigma = (\uparrow,\downarrow)$ on the $\alpha$ sublattice of $n$th unit cell. $t_\alpha$ and $\lambda_\alpha$ stand for hopping amplitude and spin-orbit coupling strength, respectively. In order to have a modulated spin-orbit coupling with uniform strength within the unit cell, we take $\lambda_{1\rightarrow T-1}\equiv \lambda$ and $\lambda_T\equiv \lambda^{\prime}$. Generalized SSH-like hopping amplitudes in the enlarged unit cells require $t_{2\alpha-1}=t-\delta t$ and $t_{2\alpha}=t+\delta t$  where $\delta t$ ($t$) is a parameter to control hopping dimerization (hopping energy). Assuming translational invariance and using periodic boundary conditions, i.e., $\hat{c}_{T+1,n,\sigma}=\hat{c}_{1,n+1,\sigma}$, we can perform Fourier transformation from $\hat{H}$ leading to $\hat{H}=\sum_{k}\hat{\psi}_k^{\dagger}\hat{h}(k)\hat{\psi}_k$ with $\hat{\psi}_k= (\hat{c}_{1,k,\downarrow},\hat{c}_{1,k,\uparrow}, ..., \hat{c}_{T,k,\downarrow},\hat{c}_{T,k,\uparrow})^\mathbf{T}$ and
\begin{equation}\label{hk-matrice}
\hat{h}(k)\!\!=\!\!\!\left(
\begin{array}{cccc}
     0 & \hat{h}_1 &  & \hat{h}_T e^{-ik} \\
     \hat{h}_1 & \ddots &  \ddots &  \\
      &  \ddots  &  \ddots & \hat{h}_{T-1} \\
     \hat{h}_T e^{ik} &   & \hat{h}_{T-1} & 0
   \end{array}
   \right)_{\!\!\!T\times T}\!\!\!\!\!\!\!\!\!.
\end{equation}
Here, we have defined $\hat{h}_\alpha=t_\alpha I+(-1)^{\alpha+1}\lambda\sigma_x$ for $\alpha=1, ...,T-1$ and $\hat{h}_T=t_T I-\lambda^{\prime}\sigma_x$ with $I$ and $\sigma_{(x,y,z)}$ being the identity matrix and $(x,y,z)$-component of Pauli matrix, respectively. The Hamiltonian (\ref{hk-matrice}) shows spatial reflection symmetry satisfying $\mathcal{R}\hat{h}(k)\mathcal{R}^{-1} = \hat{h}(-k)$ with reflection operator $\mathcal{R}=\delta_{i,T+1-j}\otimes \tau_x$ where $\delta_{i,j}$ is Kronecker delta. In addition, the system exhibits particle-hole, effective time-reversal and chiral symmetries defined, respectively, as $\mathcal{P}\hat{h}(k)\mathcal{P}^{-1} = -\hat{h}(-k)$, $\mathcal{T}\hat{h}(k)\mathcal{T}^{-1} = \hat{h}(-k)$, and $\mathcal{C}\hat{h}(k)\mathcal{C}^{-1} = -\hat{h}(k)$ whose corresponding operators are $\mathcal{P} = I_{T/2} \otimes\sigma_z\otimes\sigma_x \mathcal{K}$, $\mathcal{T} = I_{T} \otimes\sigma_x \mathcal{K}$ and $\mathcal{C} = I_{T/2}\otimes \sigma_z\otimes I$ with $\mathcal{K}$ being the complex conjugation and $I_{T}$ is an identity matrix of size $T$. Since the spatial reflection operator commutes (anti-commutes) with time-reversal (chiral) operator, the topological class of system falls into BDI with $\mathbb{Z}$ index \cite{Classification,Class1,Class2,Class3,Class4,Class5,Class6,Class7,Class8,Class9}.
The winding number as an appropriate topological invariant can be calculated in characterizing topological properties of the system possessing chiral symmetry \cite{Class10}. In this regard, we derive analytically a general formula for winding number in terms of total number of sublattices per unit cell in the following. The Hamiltonian (\ref{hk-matrice}) can be brought into block off-diagonal matrix in the basis of chiral operator to define the winding number. This can be performed through transformation $\tilde{h}_k=C\hat{h}(k)C^{-1}$  by the unitary operator $C_{T\times T}$ with nonzero matrix elements
\begin{align}\label{unitary transformation}
C_{\alpha, T-2(\alpha-1)}&=\sigma_x, &\quad 1\leq\alpha\leq T/2,&\nonumber \\
C_{\alpha, 2(T-\alpha)+1}&=\sigma_x,  &\quad T/2<\alpha\leq T, \nonumber&
\end{align}
yielding
\begin{eqnarray}
\tilde{h}(k) = \left(
                 \begin{array}{cc}
                   O & \hat{F}_k \\
                   \hat{F}^\dagger_k & O \\
                 \end{array}
               \right)
\end{eqnarray}
with
\begin{eqnarray}
\hat{F}_k =\left(
\begin{array}{cccc}
     \hat{f}_1 &O  & O & \hat{f}_2 e^{ik} \\
     \hat{f}_3 & \ddots &  O  &O   \\
     O  &  \ddots  &\ \ \ \  \ddots &O   \\
   O   & O   & \hat{f}_3 & \hat{f}_1
   \end{array}
   \right)_{\!\!T\times T}\!\!\!\!\!\!\!\!\!,
   \label{off diagonal part}
\end{eqnarray}
where $ \hat{f}_{1}=(t-\delta t) I+\lambda\sigma_x$, $\hat{f}_{2}=(t+\delta t) I-\lambda^\prime\sigma_x$ and $\hat{f}_{3}=(t+\delta t) I-\lambda\sigma_x$. Therefore, the winding number can be defined as \cite{Class10,Class11}
\begin{equation}
W =\int_{-\pi}^{\pi} \frac{dk}{2\pi i}\partial_k Ln(Det\ \hat{F}_k),
\label{winding number}
\end{equation}
where
\begin{eqnarray}
Det\ \hat{F}_k=\bigg(\gamma_1+e^{ik}\bigg)\bigg(\gamma_2-e^{ik}\bigg)\bigg(t_2^2-\lambda^{\prime^2}\bigg)\bigg(\lambda^2-t_2^2\bigg)^{\!\!\frac{T}{2}-1}\!\!\!\!\!\!\!\!,
\label{Det F}
\end{eqnarray}
and
\begin{eqnarray}
\partial_k Ln(Det\ \hat{F}_{k})=\mathcal{W}_1+\mathcal{W}_2,
\label{logdet}
\end{eqnarray}
with
\begin{align}\quad
\mathcal{W}_1&=\frac{i}{1+\gamma_1e^{-ik}},\ \ \ \ \ \ &\gamma_{1}&=\frac{(\lambda+t_1)^{\frac{T}{2}}}{(t_2-\lambda^\prime)(\lambda- t_2)^{\frac{T}{2}-1} },\\
\mathcal{W}_2&= \frac{i}{1-\gamma_2e^{-ik}},\ \ \ \ \ \ &\gamma_{2}&=\frac{(\lambda-t_1)^{\frac{T}{2}}}{(t_2+\lambda^\prime)(\lambda+ t_2)^{\frac{T}{2}-1}}.
\end{align}
Note that $Det\ \hat{F}_k$ is expressed explicitly in terms of the total number of sublattices within the unit cell. This enables us to unveil effortlessly the topological properties of the wire by enlarging the unit cell while the space parameters as well as the underlying symmetries remain intact. One can easily derive Eq. (\ref{Det F}) by analysing and comparing independently $Det\ \hat{F}_k$ for any arbitrary even number of sublattices.
The integral in Eq. (\ref{winding number}) can be evaluated using contour
integration leading to the following analytical expression
\begin{equation}
W=\begin{cases}
    2  & \quad |\gamma_1|\leq 1 \quad and \quad  \!|\gamma_2|\leq 1, \\
    1  & \quad |\gamma_1|> 1 \quad\ or \quad \ |\gamma_2|\leq 1,\\
    1  & \quad |\gamma_1|\leq 1 \quad\ or \quad \ |\gamma_2|> 1,\\
    0  & \quad |\gamma_1|>1 \quad and \quad \!|\gamma_2|>1.
  \end{cases}
  \label{analytical solution}
\end{equation}
The formula (\ref{analytical solution}) is the main result of this paper demonstrating the winding number in terms of total number of sublattices per unit cell and capturing the topological characteristic of 1D arrays easily. $W=2(1)$ indicates a nontrivial phase with fourfold (twofold) degenerate zero-energy boundary states under open boundary conditions (OBCs). $W=0$ corresponds to a trivial phase where the system is an ordinary insulator.

\section{Discussion of the results}

We write the determinant of $\hat{F}_k$ into the complex polar form, $Det\ \hat{F}_k=|Det\ \hat{F}_k|e^{i\theta_k}$. The integer $W$ enumerating the number of zero-energy boundary states under OBCs is associated with the number of times that angle $\theta_k$ winds about the origin in the complex plane. Notice this quantity is invariant under smooth perturbation and cannot change unless $|Det\ \hat{F}_k|$ goes to zero indicating a gap closing and, subsequently, topological phase transition.
Parametric plots of $Re(Det\ \hat{F}_k)$ and $Im(Det\ \hat{F}_k)$ are depicted in Fig. \ref{Energy spectrum}(b) as the momentum $k$ traverses the Brillouin zone for various values of $\lambda$ and $\lambda^{\prime}$ with $T = 4$.
The trajectories represented by solid and dotted lines wind around the origin once and twice indicating nontrivial phases corresponding to $W=1$ and $2$ resulting in twofold and fourfold degenerate zero-energy boundary states under OBCs, respectively. Also, the dashed line never encloses the origin meaning that the system is an ordinary insulator, $W=0$. Furthermore, the probability distribution for $\lambda=-2.5t$ and $2t$ with the same parameters as Fig. \ref{Energy spectrum}(b) is plotted in Fig. \ref{Energy spectrum}(c). When the dashed (solid) line with yellow (orange) arrows winds zero times (once) around the origin, the probability distribution of the lowest energy states has minimum (maxima) at the two edges of the nano-wire revealing trivial bandgap (nontrivial twofold zero-energy states). The energy spectrum versus eigenvalue index for $\delta t=t/2$ is presented in Fig. \ref{Energy spectrum}(d) showing the number of nontrivial zero-energy midgap boundary states. There is one (two) pair(s) of zero-energy midgap edge states under OBCs represented by solid orange (empty purple) circles for $\lambda=\lambda^{\prime}=2t$ ($\lambda=t/2$ and $\lambda^{\prime}=5t$) which is consistent with parametric plot of Fig. \ref{Energy spectrum}(b). Furthermore, the system is a topologically trivial band insulator for $\lambda=-2.5t$ and $\lambda^{\prime}=2t$ shown by yellow squares.
\begin{figure}[t]
\begin{center}
\includegraphics[width=8.5cm]{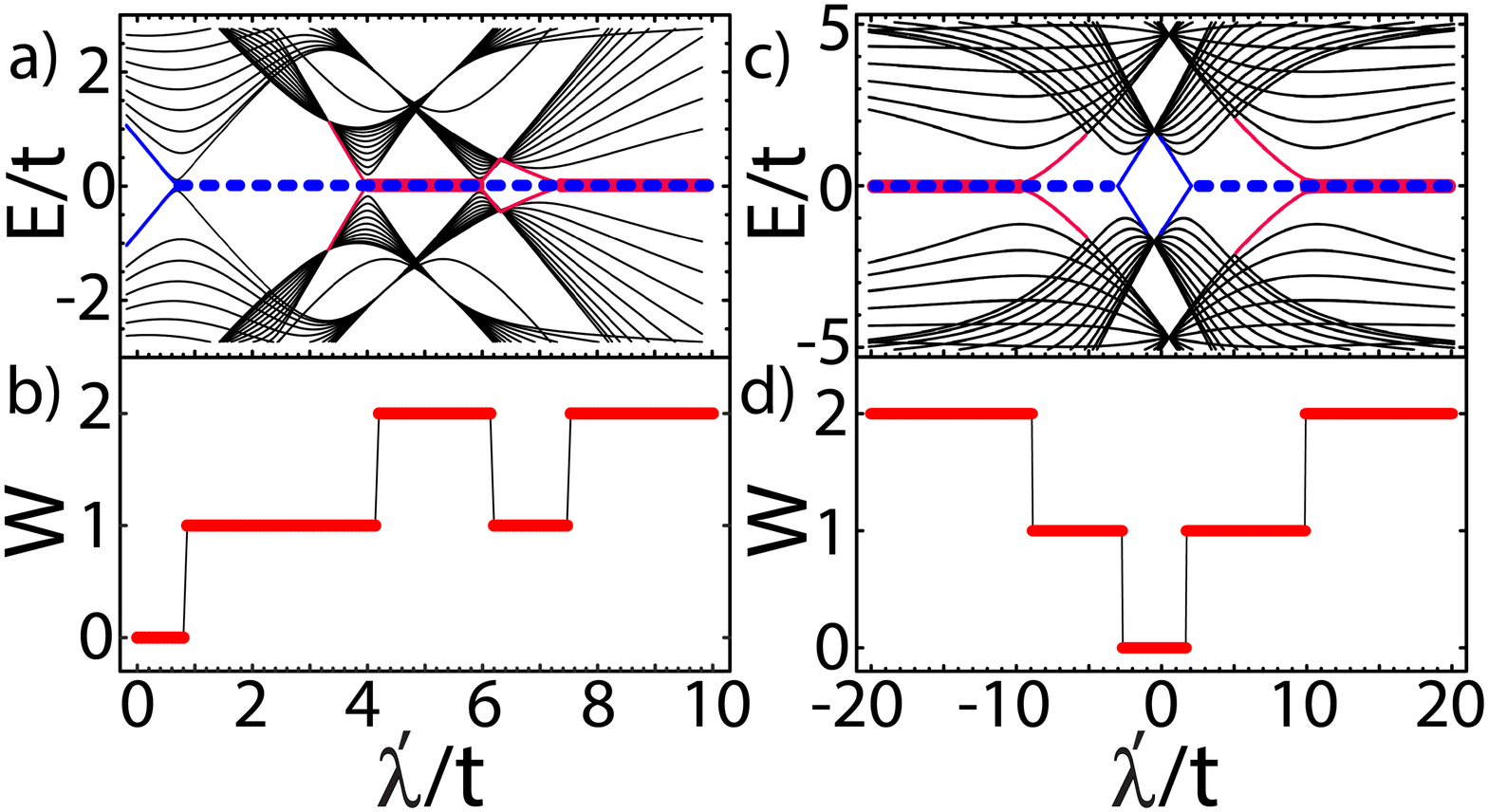}
\caption{(Color online) The energy spectrum of the finite chain as a function of $\lambda^{\prime}/t$ with (a) [(c)] $\delta t=+(-)t/2$, $N=96 (280)$ and $\lambda=-\lambda^{\prime}+5t$ ($\lambda=5t$). Bottom panels (b) and (d) are the winding numbers corresponding to the top panels.
\label{Spectra}}
\end{center}
\end{figure}

\indent The evolution of energy spectrum versus $\lambda^{\prime}$ is depicted in Fig. \ref{Spectra}(a) for $\delta t=t/2$ and $\lambda=-\lambda^{\prime}+5t$, and in Fig. \ref{Spectra}(c) for $\delta t=-t/2$ and $\lambda=5t$. In Fig. \ref{Spectra}(a), the system is a trivial insulator in small values of $\lambda^{\prime}$. As we increase $\lambda^{\prime}$, two branches of the nontrivial zero-mode boundary states (thick dashed blue line) merge together, after occurring a topological phase transition, in the parameter regime $\lambda^{\prime} \in (0.9t,4.16t)$. With further increasing of the inter cell spin-orbit coupling strength, another topological phase transition occurs at $\lambda^{\prime}=4.16t$ and bulk states touch the twofold-degenerate midgap states leading to appearance of another twofold-degenerate zero-energy boundary states (thick solid red line) in addition to the former ones. Therefore, we witness fourfold zero-energy edge states in the parameter regime $\lambda^{\prime} \in (4.16t, 6.15t)$. In the meantime, two topological phase transitions happen at $\lambda^\prime=6.15t (7.5t)$ by further increase of $\lambda^\prime$ leading to lifting (reappearing) of twofold-degenerate edge states. Moreover, Fig. \ref{Spectra}(c) has different eigenvalue structure compared to Fig. \ref{Spectra}(a), while the midgap topological states are widely similar. One can observe that zero-energy nontrivial states are available in a wide range of $\lambda^{\prime}$.  Also, the corresponding winding numbers are shown in Figs. \ref{Spectra}(b) and \ref{Spectra}(d). Evidently, as the flat bands appear with one (two) pair(s) of  boundary states eigenvalues, the winding number takes value $1 (2)$. Contrarily, if there is an energy gap, the winding number is $0$ exhibiting trivial phase where the system is an ordinary insulator.

\indent It is worthwhile mentioning that the spin-orbit couplings preserve U(1) symmetry due to rotation of spin around the lattice direction. Therefore, the appearance of fourfold-degenerate edge states highly depends on preserving the U(1) symmetry. To show this, Hamiltonian (\ref{hk-matrice}) can be brought into block diagonal in the eigenspace of spin-rotational operator $\mathcal{U}=I_{T}\otimes\sigma_x$ because of $[\hat{h}(k),\mathcal{U}]=0$ as $\hat{\mathscr{H}}({k})=\tilde{\emph{h}}_{\mathcal{U}=-}(k)\oplus\tilde{\emph{h}}_{\mathcal{U}=+}(k)$ whose decoupled subspaces are spanned by eigenstates of $\mathcal{U}$ with eigenvalues $\pm1$. This can be done through a unitary transformation $\hat{\mathscr{H}}({k})=U\hat{h}(k)U^{-1}$ where $U$ is constructed from the basis of $\mathcal{U}$ with nonzero matrix elements given by
\begin{align}\label{eigenspace of spin rotation}
U_{\alpha, T+1-\alpha}&=\hat{\mathcal{J}}, &\quad 1\leq\alpha\leq T/2,&\nonumber \\
U_{\alpha, T+1-\alpha}&=\hat{\mathcal{X}},  &\quad T/2<\alpha\leq T, \nonumber&
\end{align}
where $\hat{\mathcal{J}}=(-1,1)$ and $\hat{\mathcal{X}}=(1,1) $ are two-element row matrices. Each block of $\hat{\mathscr{H}}({k})$ takes the form
\begin{eqnarray}
\tilde{\emph{h}}_{\mathcal{U}=\pm}(k)=\left(
 \begin{array}{cccccc}
0& \Gamma^{\pm}_{1} & & & &\Gamma^{\pm}_{_{T}}e^{ik} \\
\Gamma^{\pm}_{1}& \!\!\!\!\!\!0& \Gamma^{\pm}_{2}  & & & \\
& \Gamma^{\pm}_{2}&\!\!\!\!\!\! \ddots&\ddots & &\\
& &\ddots& \ddots   &\ \ \ \ \Gamma^{\pm}_{2} & \\
 &  &  &  \Gamma^{\pm}_{2} &0 &\Gamma^{\pm}_{1} \\
 \Gamma^{\pm}_{_{T}}e^{-ik} &  &  & & \Gamma^{\pm}_{1} &\!\!\!\! 0\\
 \end{array}
    \right)_{\!\!\!\!T\times T}\!\!\!\!\!\!\!\!,
    \label{transformed Hamiltonina}
\end{eqnarray}
where $\Gamma^{\pm}_{1}=\pm\lambda+(t-\delta t)$, $\Gamma^{\pm}_{2}=\mp\lambda+(t+\delta t)$ and $\Gamma^{\pm}_{T}=\mp\lambda^\prime+(t+\delta t)$. Since each block belongs to BDI class, their band structure is gapped near the Fermi level having only twofold-degenerate zero-energy edge states in nontrivial phase. This is because the spin rotation operator commutes with the chiral, particle-hole, time-reversal, and reflection symmetries.
In consequence, by varying the parameters the zero-energy mode of an eigenspace can be overlapped with that of the other one leading to emergence of fourfold-degenerate edge states as a consequence of U(1) symmetry as illustrated in Figs. \ref{Spectra}(a) and \ref{Spectra}(c). From topological point of view, we can easily calculate the winding number of each subsystem independently. $\tilde{\emph{h}}_{\mathcal{U}=\pm}(k)$  can be transformed into block off-diagonal matrix through the unitary transformation of $\tilde{\mathscr{H}}({k})=\tilde{C}\hat{\mathscr{H}}({k})\tilde{C}^{-1}$ where $\tilde{C}$ is the basis of new chiral operator $\tilde{\mathcal{C}}$ derived in the eigenspace of spin rotational operator $\tilde{\mathcal{C}}=U\mathcal{C}U^{-1}=- I_{T} \otimes\sigma_z$ as
\begin{align}
\tilde{C}_{\alpha, T+1-\alpha}&=\hat{\mathcal{J}}^\prime, &\quad 1\leq\alpha\leq T/2,&\nonumber \\
\tilde{C}_{\alpha, T+1-\alpha}&=\hat{\mathcal{X}}^\prime,  &\quad T/2<\alpha\leq T, \nonumber&
\end{align}
and
\begin{eqnarray}
\tilde{\mathscr{H}}({k})\!\!=\!\!\left(
 \begin{array}{cccc}
 O &  O   & \hat{F}_{\mathcal{U}=+}(k) & O  \\
  O  &  O    &  O  &    \hat{F}_{\mathcal{U}=-}(k)  \\
\hat{F}^{\dagger}_{\mathcal{U}=+}(k) & O  &   O &   O \\
 O  &  \hat{F}^{\dagger}_{\mathcal{U}=-}(k)&  O  & O  \\
 \end{array}
    \!\!\!\!\right)\!,
\end{eqnarray}
with
\begin{eqnarray}
\hat{F}_{\mathcal{U}=\pm}(k)=\left(
 \begin{array}{cccc}
  \tilde{f}^{\pm}_1&  \tilde{f}^{\pm}_3  & O   & O  \\
   O &  \!\!\!\!\!\! \ddots  & \ddots &    O    \\
   O &     O    &  \ddots&\ \ \ \ \tilde{f}^{\pm}_3 \\
 \tilde{f}^{\pm}_2e^{ik}&     O   & O    &\!\!\!\!\!  \tilde{f}^{\pm}_1   \\
 \end{array}
    \right)_{\!\!\!\!T\times T}\!\!\!\!\!\!\!\!\!,
\end{eqnarray}
where $\hat{\mathcal{J}}^\prime=(1,0)$ and $\hat{\mathcal{X}}^\prime=(0,1)$, $\tilde{f}^{\pm}_1=\pm\lambda+(t-\delta t)$, $\tilde{f}^{\pm}_2=\mp\lambda^\prime+(t+\delta t)$ and $\tilde{f}^{\pm}_3=\mp\lambda+(t+\delta t)$. Utilizing the definition of winding number, Eq. (\ref{winding number}), for $\hat{F}_{\mathcal{U}=+(-)}(k)$, we interestingly get the right hand side of Eq. (\ref{logdet}) implying that $\mathcal{W}_{1(2)}$ are the portion of winding number corresponding to spin rotational eigenspaces with eigenvalues $+1(-1)$. Note the above analysis proves that U(1) symmetry leads to degeneracy of topological edges states of different subsystems.
It is worth mentioning that only y(z)-component of Zeeman magnetic field breaks the U(1) symmetry in which the subsystems couple to each other and we face with only one pair of edge states at Fermi level within the bandgap \cite{Zeeman,SOZeeman}.

\indent As already mentioned above, closing and reopening the energy gap in the system lead to topological phase transition, for which gap closure points can be obtained by setting $|Det\ \hat{F}_k|=0$. In the absence of modulated spin-orbit coupling the energy gap closes only at $k=0$ if $|t_1|=|t_2|$ and the system is topologically trivial (nontrivial) for $|t_1|<|t_2|$ ($|t_1|>|t_2|$) with none (twofold-degenerate) zero-energy midgap boundary states. In the presence of modulated spin-orbit coupling with $T$ sublattices per unit cell the energy gap closes at both $k=0$ and $k=\pm \pi$ under conditions,
\begin{eqnarray}
\lambda^\prime &=& t_2\pm\frac{(\lambda+t_1)^{\frac{T}{2}}}{(\lambda-t_2)^{\frac{T}{2}-1}},\label{topological boundary1} \\
\lambda^\prime &=& -t_2\pm\frac{(\lambda-t_1)^{\frac{T}{2}}}{(\lambda+t_2)^{\frac{T}{2}-1}}.
\label{topological boundary2}
\end{eqnarray}
\begin{figure}[t]
\begin{center}
\includegraphics[width=8.5cm]{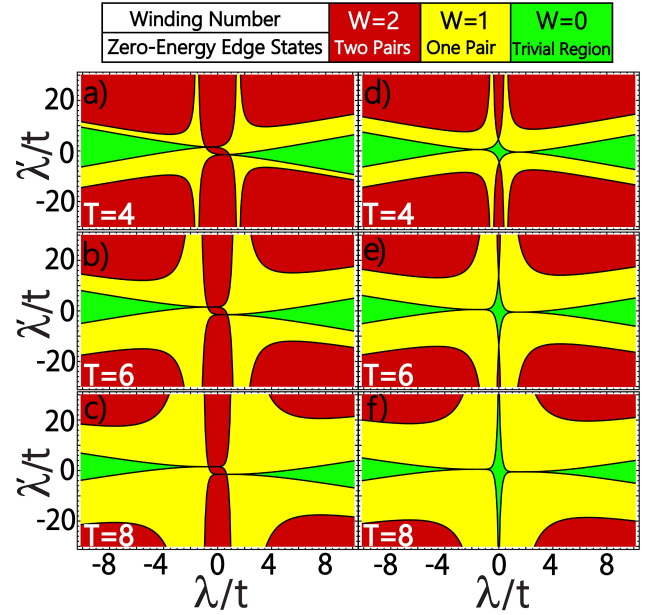}
\caption{(Color online) Topological phase diagram of the systems having even number of sublattices per unit cell in ($\lambda^{\prime}, \lambda$) plane based on the winding number calculation for $T=4$ (top panels), $T=6$ (middle panels), and $T=8$ (bottom panels). While left panels are for $\delta t=+t/2$, for right panels $\delta t=-t/2$.}
\label{Phase diagram}
\end{center}
\end{figure}
The above-obtained relations [Eqs. (\ref{topological boundary1}) and (\ref{topological boundary2})] that specify boundaries between topologically distinct phases along with Eq. (\ref{analytical solution}) can be used for determining analytically the topological phase diagrams of the system for some even number of sublattices $T$ as shown in the left and right panels of Fig. \ref{Phase diagram} for $\delta t>0$  and $\delta t<0$, respectively. $W=0 (1 or 2)$ is denoted by green color (yellow or red colors) as trivial (nontrivial) topological phase with none, (twofold or fourfold)-degenerate zero-energy edge states under OBCs. The solid lines show the topological phase transition boundaries. As it is shown in Fig. \ref{Phase diagram}(a), the unit cell contains four sublattices $T=4$ and most of the area belongs to the nontrivial ($W=1,2$) phases denoted by yellow and red colors. Thus, inter and intra cell spin-orbit couplings induce nontrivial phases in a wide range of space parameter. Clearly, with increasing the number of sublattices to $T=6$ and $T=8$ shown in Figs. \ref{Phase diagram}(b) and  \ref{Phase diagram}(c) the topological phase transition borders move in a way that $W=1(2)$ regions extend (decrease) significantly. Figure \ref{Phase diagram}(d) has a similar structure, however, the small region located at the middle of phase diagram is changed to trivial phase and the fourfold-degenerate zero-mode states are available for larger spin-orbit couplings. Similarly, enlarging the unit cell leads to increasing of regions where the twofold-degenerate zero-energy states appear as illustrated in Figs. \ref{Phase diagram}(e) and \ref{Phase diagram}(f). We also can understand that although the presence of spin-orbit coupling adds spin subspace and it mixes the spin states through spin dependent hopping, but similar to intra-cell hopping in the bare SSH model, the strong intra-cell spin-orbit coupling $\lambda$ leads to a topologically trivial phase.

\indent Finally, lets comment on the stability of the above-described topological phases. When $\lambda_{\alpha}\neq \lambda_{T-\alpha}$  the topological phases associated with the nontrivial topological flat bands of the system will be destroyed as a consequence of reflection symmetry breaking. Consequently, all the nontrivial topological phases of the system are protected by reflection symmetry and they robust against perturbations as long as such symmetry is preserved.

\section{Conclusion}

\indent We investigated the effect of spin degree of freedom in a 1D multipartite superlattice with any even number of sublattices. We showed that the spin-orbit coupling leads to appearance of nontrivial topological phases such that twofold (fourfold) degenerate zero-energy boundary states exist in the wide regions of the topological phase diagrams. We have also derived analytical formula for winding number in terms of total number of sublattices per unit cell and for gap closure points through the bulk states to characterize the topological properties of the system. Our symmetry argument shows that the appearance of fourfold-degenerate zero-energy edge states rely on the presence of U(1) spin rotational symmetry and the topological feature falls in class BDI with $\mathbb{Z}$ index as the relevant topological invariant. In this model, spin and sublattice degrees of freedom are the key factors to dominate the area of one pair of zero-energy topological edge states as the number of sublattices increases in the space of parameters. Since, the one pair of zero-energy topological edge states obeys non-Abelian statistics it can be exploited in topological quantum computation. Therefore, our model facilitates not only to reach the appropriate topological phase in systems composed of large unit cell but also to reduce the challenges of experimental preparation and manipulation of nanoscale devices in using the state-of-art technologies.

\end{document}